\documentstyle[prd,aps,epsf]{revtex}
\newcommand{\beq}{\begin{equation}}
\newcommand{\beqn}{\begin{eqnarray}}
\newcommand{\eeq}{\end{equation}}
\newcommand{\eeqn}{\end{eqnarray}}
\newcommand{\beqa}{\begin{eqnarray}}
\newcommand{\eeqa}{\end{eqnarray}}
\newcommand{\bea}{\begin{eqnarray}}
\newcommand{\eea}{\end{eqnarray}}

\def\beq{\begin{equation}}
\def\eeq{\end{equation}}

\newcommand{\square}{\kern1pt\vbox{\hrule height
1.2pt\hbox{\vrule width 1.2pt\hskip 3pt
   \vbox{\vskip 6pt}\hskip 3pt\vrule width 0.6pt}\hrule
height 0.6pt}\kern1pt}
\begin{document}

\draft \twocolumn[\hsize\textwidth\columnwidth\hsize\csname
@twocolumnfalse\endcsname

\title{Generation of electromagnetic fields in string cosmology \\
with a massive scalar field on the anti D-brane}
\author{Mohammad R.~Garousi$^1$, M. Sami$^2$ and
Shinji Tsujikawa$^3$}
\address{$^1$ Department of Physics, Ferdowsi University, P.O.Box 1436,
Mashhad, Iran\\ and\\
Institute for Studies in Theoretical Physics and Mathematics IPM,
P.O.Box 19395-5531, Tehran, Iran}
\address{$^2$ IUCAA, Post Bag 4, Ganeshkhind,
Pune 411 007, India \\[.3em]}
\address{$^3$ Department of Physics, Gunma National College of
Technology, 580 Toriba, Maebashi, Gunma 371-8530,
Japan \\[.3em]}
\date{\today}
\maketitle
\begin{abstract}
We study the generation of electromagnetic fields in a string-inspired
scenario associated with a rolling massive scalar field $\phi$
on the anti-D3 branes of KKLT  de Sitter vacua.
The 4-dimensional DBI type effective action naturally gives rise to
the coupling between the gauge fields and the inflaton $\phi$,
which leads to the production of cosmological magnetic fields
during inflation due to the breaking of conformal invariance.
We find that the amplitude of magnetic fields at decoupling epoch
can be larger than the limiting seed value required for the
galactic dynamo. We also discuss the mechanism of reheating
in our scenario and show that gauge fields are sufficiently 
enhanced for the modes deep inside 
the Hubble radius with an energy density
greater than that of the inflaton.
\end{abstract}
\vskip 1pc

\vskip 2pc
]

\underline{\em Introduction} --
The development of string theory has continuously stimulated
its application to cosmology \cite{app}-- especially to cosmic inflation.
Given the fact that the detection of some stringy effects in  accelerators
is difficult in the foreseeable future, string cosmology presumably provides
the only way to test the viability of string/M-theory
concretely. Fortunately the recent measurement of the Cosmic Microwave
Background (CMB) \cite{WMAP} has brought the high-precision
cosmological dataset by which inflationary models
can be seriously constrained  \cite{WMAPinf}.

The development in compactification with flux and branes has
provided examples of dynamics that fix all moduli \cite{GKP,KKLT}.
In particular, the authors of Ref.~\cite{KKLT} showed that by
adding anti-D3 branes to the solutions of \cite{GKP}, one can
lift the supersymmetric vacuum energy and find locally stable
minima with positive cosmological constant  (KKLT vacua).

Lately we proposed a string-inspired cosmological model \cite{GST} 
based on a massive Dirac-Born-Infeld (DBI) \cite{MRG} 
scalar field $\phi$ rolling on the anti-D3 brane of
KKLT vacua (see also Ref.~\cite{Sen}).
We showed that this scenario satisfies observational
constraints coming from the CMB temperature anisotropies by
evaluating the spectra of scalar and tensor perturbations
generated during the rolling scalar inflation \cite{GST}. 
The problem of reheating
associated with the DBI  action \cite{Frolov} is overcome by
taking into account a negative cosmological constant of the
KKLT vacua. The origin of dark energy can be
explained if the potential energy does not exactly cancel with the
negative cosmological constant at the potential minimum.

In this work we shall consider several important cosmological aspects
which distinguish our scenario from others.
It is well known that electromagnetic fields are not generated in a
Friedmann-Robertson-Walker (FRW) background
if the underlying theory is conformally invariant as in the
classical electrodynamics. On the other hand we observe
magnetic fields with magnitudes of order $10^{-6}$ G on scales
larger than 10 kpc \cite{Sofue}.
A number of people tried to solve this discrepancy
by breaking conformal invariance of the theory \cite{Seed}
or by breaking conformal flatness of the background
geometry \cite{Maroto,BPTV}.
In spite of these attempts it is fair to say that
it is not easy to construct a satisfactory cosmological model
based on particle physics which explains the origin of
seed magnetic fields and  also satisfies other observational
constraints. Presumably a unique example known earlier, 
is provided by Gasperini {\it et al.} \cite{Gas}
who highlighted the coupling between dilaton 
and gauge fields based upon the Pre-Big-Bang scenario. 
This model respects conformal
invariance which allows the photon
to decouple from the metric.

The DBI action in our string-inspired scenario involves the
coupling between gauge fields and the inflaton field $\phi$.
We show that this coupling naturally leads to the production of
cosmological magnetic fields during inflation and the amplitude
can be greater than the limiting value of the seed fields for
galactic dynamo.
We also study the reheating dynamics in which radiation is
generated very efficiently through the amplification of gauge fields.

\underline{\em Model} -- Let us consider a massive open string
excitation of the anti-D3 brane of KKLT vacua as a candidate of
the inflaton $\phi$. In addition to the massive inflaton
potential $V(\phi)$, we implement a negative
cosmological constant $(-\Lambda)$ which comes from the
stabilization of the modulus fields \cite{GKP,KKLT}. Then the DBI
type effective 4-dimensional action for our system is described by
\cite{GST}
%%%%%%%
\begin{eqnarray}
\label{action}
{\cal S} &=& \int d^4x\biggl\{\sqrt{-g}\left(M_p^2R/2
+\Lambda\right) \nonumber \\
& &-V(\phi)\sqrt{-\det(g_{\mu\nu}+
\partial_{\mu}\phi\partial_{\nu}\phi+F_{\mu \nu})}
\biggr\}\,,
\end{eqnarray}
%%%%%%%
where $V(\phi)=\beta^2T_3e^{\frac{1}{2}m^2\beta\phi^2}$,
$M_p$ is the reduced Planck mass,
and $F_{\mu \nu} \equiv 2\nabla_{[\mu}A_{\nu]}$ is
the Maxwell tensor with $A_\mu$ the four-potential.
Here $T_3$ is a brane tension, $\beta$ is a warp factor, and
$m$ is the mass of scalar vertex operators \cite{GST}.

In a spatially flat FRW metric with a scale factor $a$,
we get the background equations of motion
%%%%%%%
\begin{eqnarray}
\label{Hubble}
& &H^2 = \frac{1}{3M_p^2} \left[V(\phi)/
\sqrt{1-\dot{\phi}^2}-\Lambda \right]\,, \\
& & \ddot{\phi}/(1-\dot{\phi}^2)+3H\dot{\phi}
+\beta m^2\phi=0\,,
\label{phi}
\end{eqnarray}
%%%%%%%
where $H \equiv \dot{a}/a$ is the Hubble rate and a dot denotes
the derivative in terms of cosmic time $t$. Here we dropped
the contribution from gauge fields.
During inflation the negative cosmological constant is negligible
relative to $V(\phi)$  and inflation is realized for
$\dot{\phi}^2<2/3$. 
The presence of a negative cosmological constant
leads to a successful reheating in which the energy density of
$\phi$ scales as a pressureless dust.
In what follows we consider the generation of
electromagnetic fields during inflation and reheating.

\underline{\em Cosmological magnetic fields generated
in inflation} --
The action to the second order of gauge field becomes
%%%%%%%
\begin{eqnarray}
{\cal S}_g = \int d^4x
\frac{V(\phi)a}{2} \sqrt{1-\dot{\phi}^2}
\left(\frac{\dot{A}^2}{1-\dot{\phi}^2}-\frac{
\vec{\nabla}A\cdot\vec{{\nabla}}A}{a^2}\right),
\end{eqnarray}
%%%%%%%
where we used the coulomb gauge
($A_0=0=\vec{\nabla}\cdot\vec{A}$), and $ \vec{A}\equiv A$.
{}From this Lagrangian we get the equation for
the each Fourier component of the gauge field
%%%%%%%
\begin{eqnarray}
\label{Ak}
\ddot{A_k}+H(1-3\dot{\phi}^2)\dot{A_k}+(1-\dot{\phi}^2)
(k^2/a^2)A_k=0\,,
\end{eqnarray}
%%%%%%%
where $k$ is a comoving wavenumber.
Introducing a new quantity $\tilde{A}_k=b^{1/2}A_k$
with $b=V(\phi)/\sqrt{1-\dot{\phi}^2}$, we find
%%%%%%%
\begin{eqnarray}
\label{Ak2}
\tilde{A}_k''+ \left[k^2-f(\eta)\right]
\tilde{A}_k=0\,,
\end{eqnarray}
%%%%%%%
where a prime denotes the derivative with respect to
a conformal time $\eta=\int a^{-1}{\rm d}t$ and
%%%%%%%
\begin{eqnarray}
\label{fre}
f(\eta)=k^2\dot{\phi}^2-3a^2\dot{\phi}
\left(H\ddot{\phi}+\frac{H^2\dot{\phi}}{2}
+\frac{\dot{H}\dot{\phi}}{2}-\frac{3H^2
\dot{\phi}^3}{4}\right).
\end{eqnarray}
%%%%%%%

In an asymptotic past ($\eta \to -\infty$) with $\dot{\phi} \to 0$,
$f(\eta)$ is vanishingly small compared to
$k^2$ and the solution is given by
$\tilde{A}_k^{i}=e^{-ik\eta}/\sqrt{2k}$.
The $\dot{\phi}$ term gradually becomes important
during inflation, which leads to the variation of
$\tilde{A}_k$.
The equation (\ref{Ak2}) has the following  solution:
%%%%%%%
\begin{eqnarray}
\label{bogo}
\tilde{A}_k=\alpha_k \tilde{A}_k^{i}+
\beta_k \tilde{A}_k^{i *}\,,
\end{eqnarray}
%%%%%%%
where the Bogolyubov coefficient $\beta_k$ is
%%%%%%%
\begin{eqnarray}
\label{alphabeta}
\beta_k = -i \int_{\eta_i}^{\eta}
\tilde{A}_k^i f(\eta') \tilde{A}_k {\rm d} \eta'\,.
\label{beta}
\end{eqnarray}
%%%%%%%

\begin{figure}
\epsfxsize = 3.3in \epsffile{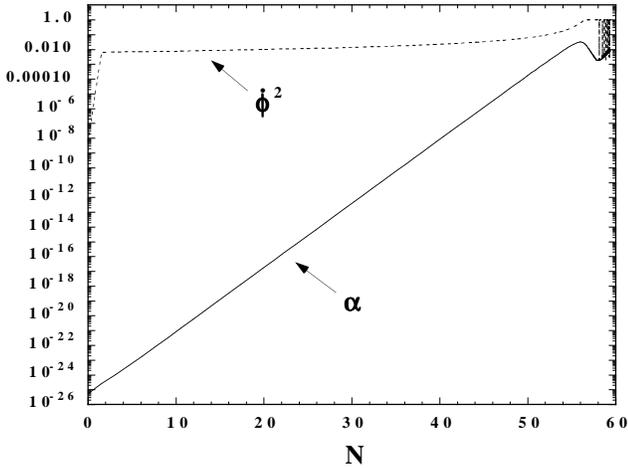}
\caption{The evolution of $\alpha$ and $\dot{\phi}^2$
during inflation for $\beta=10^{-9}$ as a function of the $e$-folds
$N=\int H{\rm d}t$.
We choose the cosmological mode which crossed the
Hubble radius about 56 $e$-folds before the end of inflation.
We find that
$\alpha$ reaches of order 0.01 at the end of inflation.}
\label{alpha}
\end{figure}

The spectrum of the gauge field $A_k$ can be evaluated analytically
by using the method in Ref.~\cite{Dimo}.
After the Hubble radius crossing ($k \lesssim aH$), the mass term 
in Eq.~(\ref{Ak2}) is approximately given by 
$-f(\eta)\simeq -k^2\dot{\phi}^2+(3/2)\dot{\phi}^2(aH)^2 \sim 
\dot{\phi}^2/\eta^2$ under a slow-roll approximation.
We shall consider a situation in which the field $\phi$
rapidly decays to radiation after inflation and
the gauge field becomes effectively massless during the 
subsequent radiation dominant era.  
Then one can find the spectrum of super-Hubble gauge fields
by matching two solutions
between inflation and radiation (see sec.~4.2 of \cite{Dimo}). 
This gives the spectrum ${\cal P}_{\tilde{A}_k} \equiv k^3/(2\pi^2)
|\tilde{A}_k^2| \propto k^n$ with $n$ of order $\dot{\phi}^2$ 
[see Eq.~(90) of \cite{Dimo}]. Since $\dot{\phi^2}$
is much smaller than 1 (see Fig.~\ref{alpha}), the resulting 
spectrum is nearly scale-invariant. 
In the case of 
$Z$-boson with mass $M_Z$ the spectral index is 
$n=2(M_Z/H)^2$ \cite{Dimo}.
There exists an interesting mechanism to amplify magnetic fields by parametric 
resonance during reheating using a coupling between a gauge field and 
a Klein-Gordon scalar \cite{finelli}. In this case the effective mass 
of gauge fields during inflation is typically required to be larger than 
$H$ for parametric resonance to be efficient, thereby giving a blue-spectrum
$n>1$.

The energy density stored in a magnetic field mode
$|B_k|$ is given by \cite{BPTV}
%%%%%%%
\begin{eqnarray}
\label{rhoB}
\rho_B=|B_k|^2/(8\pi)=\omega^4 |\beta_k|^2\,,
\end{eqnarray}
%%%%%%%
where $\omega=k/a$. In the presence of the $f(\eta)$ term,
$|\beta_k|^2$ is not generally zero, which corresponds
the generation of electromagnetic fields.
Making use of Eqs.~(\ref{beta}) and (\ref{rhoB})
the amplitude of the magnetic field is expressed as
%%%%%%%
\begin{eqnarray}
|B_k|=\sqrt{8\pi}\frac{k^2}{a^2} |\beta_k|=
\sqrt{2\pi}\frac{k}{a^2} \left| \int_{\eta_i}^{\eta_f}
Q_k^i f(\eta) Q_k {\rm d}\eta \right|\,,
\end{eqnarray}
%%%%%%%
where we introduced a dimensionless quantity
$Q_k=\sqrt{2k}\tilde{A}_k$.
Note that $\eta_f$ is the time after which the particle creation
ceases. We are interested in the amplitude $|B_k^{dec}|$
on a scale corresponding to the time at decoupling,
i.e., $\omega_{dec}=
k_{dec}/a_{dec} \simeq 10^{-33}\,{\rm GeV}$.
{}From the above equation we obtain
%%%%%%%
\begin{eqnarray}
\label{Bdec}
|B_k^{dec}|=\sqrt{2\pi}\omega_{dec}H_e
(a_e/a_{dec}) \alpha \,,
\end{eqnarray}
%%%%%%%
where $\alpha$ is a dimensionless quantity, given by
%%%%%%%
\begin{eqnarray}
\alpha=\left|\int_{\eta_i}^{\eta_f}
Q_k^i g(\eta) Q_k \frac{(aH)^2}{a_eH_e}
{\rm d}\eta \right| \,.
\end{eqnarray}
%%%%%%%
Here $g(\eta)=(aH)^{-2}f(\eta)$ and the subscript
`{\it e}' denotes the value at the end of inflation.

Let us estimate the amplitude of cosmological magnetic fields
generated in our scenario. The slow-roll parameter $\epsilon
=-\dot{H}/H^2$ becomes greater than 1 when $\dot{\phi}^2$ reaches
$\dot{\phi}^2 \simeq 2/3$, from which one can find the value of
$\phi$ at the end of inflation as $\phi_e=6.715/(\sqrt{\beta}m)$
by using Eqs.~(\ref{Hubble}) and (\ref{phi}). The COBE
normalization gives $\beta \simeq 10^{-9}$ for an exponential
potential \cite{GST}, thereby yielding $H_e \simeq 6.0 \times
10^{-5}M_p \simeq 10^{14}$ GeV. 
%It is interesting to note that
%the value of inflaton during the inflation epoch is much larger
%that the radius of the compact space which is of order $1/m$.

%%%
%%%

The ratio $a_e/a_{dec}$ depends on the details of reheating.
For example if the energy density of the field $\phi$ at the end of
inflation is converted to almost instantaneously to radiation,
$\rho_R=(\pi^2/30)g_*T_R^4$, where $g_* \sim 100$ is the number
of relativistic degree of freedom, then we obtain the reheating
temperature $T_R \sim 10^{15}$ GeV for $\beta=10^{-9}$, which
yields $a_e/a_{dec} \sim T_{dec}/T_R \sim 10^{-25}$. When the
thermalization process is delayed, one gets the smaller reheating
temperature, thereby giving larger values of $T_{dec}/T_R$.
By Eq.~(\ref{Bdec}) the amplitude of the cosmological
magnetic field at decoupling epoch is estimated as
%%%%%%%
\begin{eqnarray}
\left|B_k^{dec}\right| \simeq (a_e/a_{dec})
\alpha\,~{\rm G}\,,
\end{eqnarray}
%%%%%%%
where we used $1\,{\rm G}^2/8\pi=
1.9089 \times 10^{-40}\,{\rm GeV}^4$.

The size of $\alpha$ acquired during inflation
is crucially important to estimate the amplitude of
magnetic fields.
As shown in Fig.~\ref{alpha}, $\alpha$ rapidly grows during
the inflationary stage. This comes from the fact that
the second term in Eq.~(\ref{fre}) increases
due to the presence of the $a^2$ term,
which becomes dominant after the Hubble radius crossing
even if the $\dot{\phi}^2$ term is suppressed during inflation.
We choose the initial condition $\dot{\phi}_i=0$
in our numerical simulations, but
$\dot{\phi}^2$ is of order 0.01 in most stage of inflation
as seen in Fig.~\ref{alpha}.
The evolution of $\alpha$ becomes
mild during the reheating phase and
the particle creation ends
when the system approaches another asymptotic limit
$\eta_f \to \infty$ characterized by $\dot{\phi} \to 0$.

In the case of instant reheating with
$\beta=10^{-9}$, one has $a_e/a_{dec} \sim 10^{-25}$,
yielding $|B_k^{dec}| \sim 10^{-27}$ G for $\alpha=0.01$.
It is known that seed fields with the amplitude
larger than $|B_k^{dec}|=10^{-23}$ G
is required for the galactic dynamo for a flat universe without a cosmological
constant, but this limit is relaxed up to $|B_k^{dec}|=10^{-30}$ G
in the presence of a cosmological constant with
$\Omega_\Lambda=0.7$ \cite{Davis}.
Therefore the value $|B_k^{dec}| \sim 10^{-27}$ G is greater than
this limiting value.
Thus our scenario provides a sound mechanism to
generate seeds magnetic fields through the breaking of
conformal invariance.
Larger magnetic fields can be obtained if the ratio $a_e/a_{dec}$
is higher and this depends on the details of reheating.

\underline{\em Reheating from electromagnetic fields} --
Let us next consider how the gauge fields are produced in reheating stage
after inflation. We denote the comoving wavenumber corresponding
to the Hubble radius at the end of inflation as $k=a_eH_e$.
Then the energy density of produced gauge fields
can be expressed as
%%%%%%%
\begin{eqnarray}
\label{rhordef}
\rho_B(k)=\left(\frac{k}{a_eH_e}\right)^4
\left(\frac{a_e}{a}\right)^4 |\beta_k|^2 H_e^4\,.
\end{eqnarray}
%%%%%%%

One can estimate the size of gauge fields generated
during {\it inflation} for the mode $k \sim a_eH_e$.
Since the frequency of $\tilde{A}_k$ is approximately given by
$\omega_k^2 \simeq k^2(1-\dot{\phi}^2)$
for these modes during inflation, the Bogolyubov coefficient
$\beta_k$ at the end of inflation is estimated as
%%%%%%%
\begin{eqnarray}
\beta_k \simeq (1/4) \langle \dot{\phi}^2 \rangle
\left(e^{-2ik\eta_e}-e^{-2ik\eta_i}\right)\,,
\end{eqnarray}
%%%%%%%
where we used $\tilde{A}_k \simeq \tilde{A}_k^i$.
Note that $\langle \dot{\phi}^2 \rangle$ is the average value
of $\dot{\phi}^2$, which means that $|\beta_k|^2 \ll 1$
at the end of inflation for the mode $k \sim a_eH_e$.
Therefore $\rho_B(k)$ is much smaller than 
the energy density of the field $\phi$ ($\rho_\phi$)
at the beginning of reheating. 

When the reheating stage starts, the $\dot{\phi}^2$ term
rapidly grows to unity as seen in Fig.~\ref{alpha}.
We numerically found that $\dot{\phi}$
oscillates between $-1$ and $1$ with a slow
adiabatic damping due to cosmic expansion.
During the oscillation, $\dot{\phi}$ tends
to stay around $\dot{\phi} \simeq \pm 1$ for some moment
of time as is checked from Eq.~(\ref{phi}).
While this oscillation is not sinusoidal,
one may expect that gauge fields are generated by
parametric resonance due to the non-adiabatic change of
the frequency in $\tilde{A}_k$.

Actually this happens depending on the size of the momentum
$k$. The gauge fields $\tilde{A}_k$ for the modes deep inside the
Hubble radius ($k \gg a_eH_e$) exhibit rapid growth,
whereas the modes corresponding to $k \lesssim a_eH_e$
are not amplified. This can be understood as follows.
Introducing a new quantity, $\bar{A}_k=a^{1/2}\tilde{A}_k$,
the equation for gauge fields for the mode $k \gg a_eH_e$
is approximately given as
%%%%%%%
\begin{eqnarray}
\label{hatAk}
\ddot{\bar{A}}_k+(k^2/a^2)
(1-\dot{\phi}^2)\bar{A}_k \simeq 0\,,
\end{eqnarray}
%%%%%%%
which has an oscillating term $\dot{\phi}^2$.
The approximate time scale for the oscillation of the field $\phi$
is $m_{\rm eff}^{-1}$ with $m_{\rm eff} \equiv \sqrt{\beta}m$.
We shall define the effective resonance parameter
%%%%%%%
\begin{eqnarray}
q \equiv \frac{k^2}{4a^2m_{\rm eff}^2}=
\frac14 \left(\frac{k}{a_eH_e}\right)^2
\left(\frac{a_e}{a}\right)^2 \left(\frac{H_e}{m_{\rm eff}}
\right)^2\,,
\end{eqnarray}
%%%%%%%
which is analogous to the one defined in Ref.~\cite{KLS}.

\begin{figure}
\epsfxsize = 3.3in \epsffile{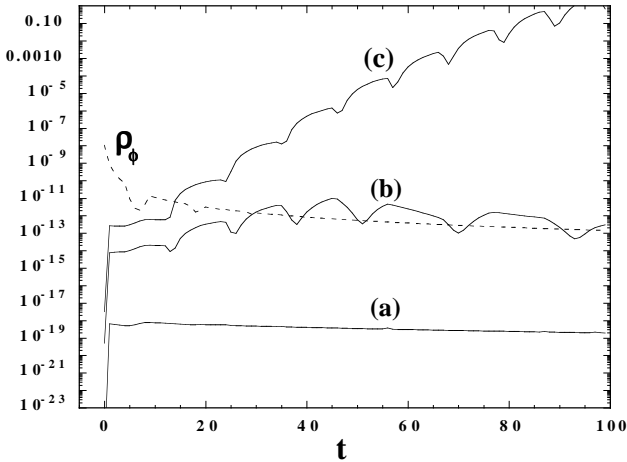}
\caption{The evolution of $\rho_B(k)$
during reheating with $\beta=10^{-9}$ for
(a) $k/(a_eH_e)=1$, (b) $k/(a_eH_e)=10$
and (c) $k/(a_eH_e)=20$.
We also show the evolution of the energy density
for the field $\phi$ (denoted by $\rho_\phi$).
We find that $\rho_B(k)$ surpasses $\rho_\phi$
for the mode $k/(a_eH_e) >10$.
}
\label{rhor}
\end{figure}

The parameter $q$ characterises the strength of parametric
resonance. The modes deep inside the Hubble radius
($k \gg a_eH_e$) correspond to the large resonance parameter
($q \gg 1$) at the of inflation (here we used the fact that
$H_e$ is the same order as $m_{\rm eff}$).
As shown in  Fig.~\ref{rhor}
we find that the production of gauge fields for the
modes $k \gtrsim 10a_eH_e$  is so efficient that
its energy density can surpass that of the field $\phi$.
Note that there is no 
amplification of the fluctuation $\delta\phi$ in our 
scenario at the linear level \cite{GST}.
The excitation of gauge fields continues until
the coherent oscillation of the
field $\phi$ is broken by the backreaction effect of produced
particles or the parameter $q$ drops down to of order unity.
For the modes $k \lesssim a_eH_e$ the resonance parameter
$q$ is smaller than unity, which means that the
particle creation is not efficient as seen in Fig.~\ref{rhor}.
The cosmological magnetic field is involved
in this case ($k \ll a_eH_e$), implying that
its generation is weak during reheating.

%%%
%%%

In Fig \ref{rhor} we do not account for backreaction 
and rescattering effects of created particles.
This is particularly important
when $\rho_B(k)$ catches up $\rho_\phi$.
In addition the effect of conductivity is not negligible
if charged particles are produced after inflation.
This typically suppresses the growth of gauge fields \cite{BPTV},
whose effect can be important after the intermediate stage of reheating.
While the explosive production of gauge fields at the initial
stage  is robust, we require a fully nonlinear analysis
including the effect of conductivity in order to understand the
reheating dynamics completely.
The process of the thermalization is also dependent on the details of
such a nonlinear stage. It is of interest to do a detailed analysis in such
a regime, since the reheating temperature is relevant to estimate the
amplitude of cosmological magnetic fields.

\underline{\em Conclusions} --
Generally it is difficult to construct a viable cosmological model
in string theory which satisfies all cosmological/observational
constraints including inflation, reheating, dark energy and
primordial magnetic fields.
However our scenario using a massive rolling scalar on the
anti-D3 brane of KKLT vacua can provide a satisfactory explanation
for the above requirements.

The crucial point for the generation of cosmological magnetic
fields in our model is that the DBI type action breaks 
conformal invariance due to the coupling between gauge fields
and the inflaton. This also happens for the rolling tachyon field
\cite{tachyon}, implying that magnetic fields can be generated
during tachyon inflation. Nevertheless tachyon inflation
generally suffers from the problems associated with large density
perturbations and reheating \cite{Frolov}. In our scenario these
problems are overcome by considering a warp metric with a small
parameter $\beta (\ll 1)$ and also by a negative cosmological
constant, both appearing in the KKLT vacua \cite{KKLT}.

During reheating we find that radiation is efficiently produced
by the amplification of gauge fields on sub-Hubble scales.
This corresponds to the ``preheating'' stage as in the bosonic
particle creation through the coupling
$(1/2)g^2\phi^2\chi^2$ \cite{KLS}. The important point is that
we used only the DBI type string effective action without adding
any phenomenological terms.  It is of interest to extend
our analysis to the production of fermions using the
couplings appearing in string theory.

%%%%%%%%%%%%%%%%%

\end{document}